\documentclass[
 reprint,
 superscriptaddress,
 amsmath,amssymb,
 aps,
 pra,
]{revtex4-2}

\usepackage{graphicx}
\usepackage{dcolumn}
\usepackage{bm}

\usepackage[version=4]{mhchem}
\usepackage[utf8]{inputenc}
\usepackage[T1]{fontenc}

\usepackage{hyperref}

\usepackage{booktabs}

\usepackage{soul}
\usepackage[dvipsnames]{xcolor}

\newcommand{\nspins}{{n}}

\begin{document}

\preprint{APS/123-QED}
 
\title[Predicting sampling advantage of stochastic Ising Machines for Quantum Simulations]{Predicting sampling advantage of stochastic Ising Machines for Quantum Simulations}

\author{Rutger J.L.F. Berns}
\email{Contact author: rutger.berns@ru.nl}
\affiliation{ 
Radboud University, Institute for Molecules and Materials, 6525 AJ Nijmegen, Heyendaalseweg 135, The Netherlands
}
\author{Davi R. Rodrigues}
\affiliation{Department of Electrical and Information Engineering, Politecnico di Bari, 70126, Bari, Italy}
\author{Giovanni Finocchio}
\affiliation{%
Department of Mathematical and Computer Sciences, Physical Sciences and Earth Sciences, University of
Messina, 98166, Messina, Italy
}
\author{Johan H. Mentink}
\affiliation{ 
Radboud University, Institute for Molecules and Materials, 6525 AJ Nijmegen, Heyendaalseweg 135, The Netherlands
}

\date{\today}

  \begin{abstract}
    
    Stochastic Ising machines, sIMs, are highly promising accelerators for
    optimization and sampling of computational problems that can be formulated
    as an Ising model. Here we investigate the computational advantage of sIM
    for simulations of quantum magnets with neural-network quantum states (NQS),
    in which the quantum many-body wave function is mapped onto an Ising model.
    We study the sampling performance of sIM for NQS by comparing sampling on a
    software-emulated sIM with standard Metropolis-Hastings sampling for NQS. We
    quantify the sampling efficiency by the number of {computational}
    steps required to reach iso-accurate stochastic estimation of the
    variational energy and show that this is entirely determined by the
    autocorrelation time of the sampling. This enables {predictions} of
    sampling advantage without direct deployment on hardware. {Although
    sampling of the quantum Heisenberg models studied exhibits much longer
    autocorrelation times on sIMs, the massively parallel sampling of hardware
    sIMs leads to a projected} speed-up of 100 to 10000, suggesting great
    opportunities for studying complex quantum systems at larger scales.
\end{abstract}

\maketitle

\section{\label{sec:intro}Introduction} Stochastic Ising Machines (sIM) are an
unconventional computational paradigm with high potential to solve certain
NP-hard problems, like combinatorial optimization problems, faster and for large
problem sizes
\cite{mohseni_isingmachineshardware_2022,aadit_massivelyparallelprobabilistic_2022}.
To this end, the problem is mapped first onto an Ising model, which is then
implemented as physical hardware which solves the Ising model in a massively
parallel and energy-efficient way. A particularly promising technique to realize
sIMs is using probabilistic bits (p-bits), which can fluctuate between 0 and 1
\cite{camsari_pbitsprobabilisticspin_2019,
pervaiz_hardwareemulationstochastic_2017}. The advantage of p-bits is that they
can be implemented on large scale and operate massively parallel with up to
millions of p-bits on a single chip, realizing up to petaflips per second
\cite{sutton_autonomousprobabilisticcoprocessing_2020}.

Current demonstrations with sIMs mostly focus on benchmarking canonical
combinatorial optimization problems such as SAT and Max-CUT
\cite{aadit_massivelyparallelprobabilistic_2022,mcmahon_fullyprogrammable100spin_2016}.
Another promising application area is quantum simulation
\cite{chowdhury_fullstackviewprobabilistic_2023}, for example quantum Monte
Carlo (QMC) for simulating ground state problems and the recently introduced
Neural Network Quantum States (NQS) \cite{carleo_solvingquantummanybody_2017},
which additionally can simulate unitary quantum dynamics
\cite{fabiani_investigatingultrafastquantum_2019,schmitt_quantummanybodydynamics_2020,
reh_optimizingdesignchoices_2023, sinibaldi_timedependentneuralgalerkin_2024}.
However, it is difficult to predict how benchmarks on SAT and Max-CUT problems
translate to a potential computational advantage for probabilistic simulation of
such quantum problems. 

In both QMC and NQS, the quantum problem in dimension $D$ can be constructed as
a mapping onto a classical Ising model of dimension $D^\prime > D$ on which
importance sampling is performed, which in principle is very well suited for
acceleration on p-bit computers. For the NQS use case considered below,
restricted (RBM) and deep Boltzmann machines (DBM) are typically considered as
variational ansatze for the quantum many-body wavefunction
\cite{carleo_solvingquantummanybody_2017,carleo_constructingexactrepresentations_2018}.
These improve upon traditional ansatze \cite{carleo_solvingquantummanybody_2017}
and led to new discoveries
\cite{fabiani_supermagnonicpropagationtwodimensional_2021,
nomura_diractypenodalspin_2021}. Estimating observables, such as the variational
energy, results in a computational complexity that scales quadratically with the
number of spins, currently limiting simulations of systems to about 1000 quantum
spins \cite{fabiani_investigatingultrafastquantum_2019,
moss_leveragingrecurrenceneural_2025}. Moreover, the simulation of DBMs, which
fundamentally can represent any quantum state efficiently, possess even higher
sampling complexity yet may lead to extremely high-fidelity quantum simulations
\cite{nomura_purifyingdeepboltzmann_2021,
nomura_boltzmannmachinesquantum_2024,carleo_constructingexactrepresentations_2018,deng_quantumentanglementneural_2017}.
Therefore, it is very interesting to find scaling advantages of sIMs for NQS. 

Given the large range of hardware implementations of sIMs
\cite{mohseni_isingmachineshardware_2022}, directly benchmarking is challenging
even for a single use case and a given NQS architecture. Therefore, in this work
we focus on an alternative method to identify potential computational advantage
of sIMs for NQS, focusing on the number of {computational} steps that a
given sampling algorithm needs on both existing digital and reported designs of
p-bit computers. We show that the number of steps can be accurately predicted
based on the autocorrelation time of the sampling algorithm. We compare both the
Metropolis-Hastings (MH) algorithm and Gibbs sampling, which are the workhorse
for traditional NQS and sIM implementations, respectively. This approach has the
advantage that it provides a hardware agnostic measure of the runtime required
that is only defined by the sampling efficiency of the Ising model used to
represent quantum states. Evaluation of this approach to pre-trained RBM models
and combining the results with reported runtime per step, we find 2 to 4 orders
of magnitude speed up compared to single chain MH sampling.

The remainder of this paper is organized as follows. First, we discuss the
quantum Heisenberg model used to investigate sampling advantage and the mapping
of neural network quantum states to a sIM. Second, we present a method to
determine the number of steps required on a sIM compared to MH sampling to
obtain the same accuracy, which we evaluate from sampling of pre-trained RBM
models. Third, we evaluate the results of this method for the ground state
energy of the quantum Heisenberg model, use this to predict sampling advantages
and discuss the implications of the results. Finally, we discuss the results and
draw conclusions.

\section{\label{sec:methods} Model and method}

The quantum Heisenberg model is an effective model that describes the
interaction between nearest-neighbor spins on a lattice and is described by the
following Hamiltonian:
\begin{equation}
    \hat{H} =  J_\text{ex} \sum_{\langle i j\rangle} \hat{\mathbf{S}}_i \cdot \hat{\mathbf{S}}_j 
    \label{eq:hamiltonian:heisenberg}
\end{equation}
where $J_\text{ex}$ is the exchange interaction between spins at neighboring
lattice sites and the summation runs over all such pairs of lattice sites. We
consider a square lattice with periodic boundary conditions and the
$\hat{\mathbf{S}}_i$ are the quantum-mechanical spin operators on the site $i$,
for $S=1/2$. Our main interest will be in the antiferromagnetic case
($J_\text{ex}>0$), which is widely studied owing to the relevance of cuprates
used for high-temperature superconductors
\cite{lorenzana_doesheisenbergmodel_1999,
manousakis_spin1/2heisenbergantiferromagnet_1991} and has also been extensively
used for benchmarks of NQS
\cite{carleo_solvingquantummanybody_2017,fabiani_investigatingultrafastquantum_2019,
wu_variationalbenchmarksquantum_2024}.

\begin{figure}
    \centering
    \includegraphics{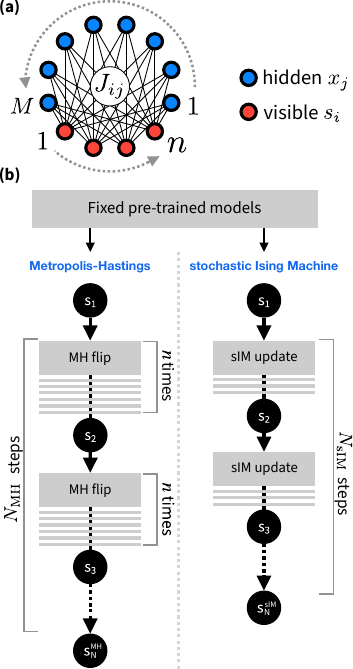}
    \caption{\label{fig:figuremethod} \textbf{(a)} Restricted Boltzmann Machine
    with $n$ visible spins and ${m}$ hidden spins. The visible spins
    (red) are fully connected to the hidden spins (blue), but there are no
    intralayer connections. The connections correspond to non-zero weights in
    the weight matrix $J_{ij}$. \textbf{(b)} Setup of the sampling methods. The
    MH sampling on the left performs of $N_\mathrm{MH}$ MH sweeps each
    consisting of $n$ antiparallel spin flips of the visible spins. The
    stochastic Ising Machine (sIM) on the right performs
    {$N_\mathrm{sIM}$ sIM sweeps which corresponds to the number of
    samples in the magnetization zero sector.}}
    \end{figure}

For concreteness, the focus of our study will be the ground state energy of the
2D quantum Heisenberg model which we study on the basis of NQS. The ground state
wavefunction of a quantum system is a complex probability distribution. The RBM,
Fig. \ref{fig:figuremethod}a, is a powerful neural network to model such
probability distributions
\cite{leroux_representationalpowerrestricted_2008,carleo_solvingquantummanybody_2017}
and consists of two types of spins, the visible corresponding to the physical
quantum spins and the hidden spins which act as auxiliary variables. The visible
spins are fully connected to the hidden spins and there are no intra-layer
connections. The probability $P(s) = |\psi(s)|^2$ to observe a state $s$ is
given by
\begin{equation}
    P (s) = \sum_{\{x_j\}} e^{\sum_{j=1}^{{m}} b_j x_j + \sum_{i=1,j=1}^{n, {m}} W_{ij} s_i x_j}
    \label{ising:equation:rbmprobability}
\end{equation}
where $s_i$ refers to the z-projection of physical/visible spins, $b_j$ are the
real biases of the hidden spins $x_j$ and $W_{ij}$ is the real weight matrix
connecting the visible and hidden. In total there are $n$ visible spins and
${m} = \alpha n$ hidden spins, where $\alpha$ is also known as the
hidden layer density. Increasing $\alpha$ increases the expressiveness of the
RBM.

The wave function ansatz is then given by
\begin{equation}
    \psi(s) = \exp \left[\frac{1}{2} \sum_{{j}=1}^{{m}} \log{ \left( 2 \cosh{\left(b_{{j}} + \sum_{{i=1}}^{{n}} W_{ij} s_{{i}} \right)}\right) }\right].
    \label{ising:equation:modifiedansatz}
\end{equation}
where the hidden spins $\{x_{{j}}\}$ are traced out. For general
Hamiltonians $\psi(s)$ needs to be complex, but for the Heisenberg model
$\psi(s)$ can be chosen real \cite{marshall_antiferromagnetism_1955} also
allowing real parameters. 

We are interested in extracting samples $s$ that are distributed according to
$|\psi(s)|^2$, {using an equivalent Ising Model of which the
Hamiltonian is given
by}
\begin{equation}
    H_\text{ising}({z}) = -\sum_{i = 1}^{{m}+n} h_i {z}_i - \sum_{j=1}^{{m}+n} \sum_{i=1}^{j-1} J_{ij} {z}_i {z}_j
    \label{ising:equation:isinghamiltonian}
\end{equation}
and ${P(z) = e^{-H_\text{ising}(z)}}$, where the spins are encoded in a single
vector ${z}= [x_{1} x_2  \dots x_{{m}}  \ s_1 \ s_2 \dots  s_{n}]$.
{Note that the connectivity of the Ising Hamiltonian here is
all-to-all, which differs from the
canonical nearest-neighbor Ising model typically used in physics}. The weight
matrix $\mathbf{J}$ is given by
\begin{equation}
\mathbf{J} = 
\begin{bmatrix}
    \mathbf{0} & (\mathbf{W}^T)_{{m \times n}} \\
    \mathbf{W}_{{n \times m}} & \mathbf{0}
\end{bmatrix}
\label{ising:equation:Jmatrix}
\end{equation}
and the bias term $\mathbf{h}$ is given by
\begin{equation}
    \mathbf{h} = [\mathbf{b}_{{m}} \ \mathbf{0}_{{n}}].
    \label{ising:equation:biasvector}
\end{equation}

Our main interest is the variational estimation of the ground state energy,
which is given by
\begin{equation}
    E = \frac{\langle \psi | \hat{H} | \psi \rangle}{\langle \psi | \psi \rangle} = \sum_{s}P(s) E_\mathrm{loc}(s) \approx \frac{1}{N} \sum_{{k}=1}^{{M}} E_\mathrm{loc}(s_{{k}})
    \label{eq:energy}
\end{equation}
where $M$ is the number of samples and $E_\mathrm{loc}(s)$ is given by
\begin{equation}
    E_\mathrm{loc}(s) = \frac{\langle s | \hat{H} | \psi \rangle}{\langle s | \psi \rangle}.
    \label{eq:local_energy}
\end{equation}
The states $s_{{k}}$ are either obtained by MH sampling on the visible
spins or sampling the introduced Ising model on an (emulated) sIM, see Fig.
\ref{fig:figuremethod}b.

The sIM samples both hidden and visible spins and uses Gibbs sampling. A key
advantage of the RBM's bipartite structure is that it enables the parallel
updating of unconnected spins, namely the visible and hidden, see Fig.
\ref{fig:figuremethod}a. This method, known as chromatic or blocking Gibbs
sampling
\cite{gonzalez_parallelgibbssampling_2011,aarts_simulatedannealingboltzmann_1989},
allows for intrinsic parallelism, making it particularly well-suited for
deployment on physical hardware.

On the other hand, in the MH sampling traditionally used for NQS, only the
visible spins are sampled. Additionally, the MH sampling uses local updates,
like a pair of antiparallel spins, which conserves the total
magnetization{, $\sum_{i=1}^n s_i$}. This is convenient because
sampling can be limited to the magnetization zero sector since the total spin
operator $\hat{S}_\mathrm{tot} = \sum_{i=1}^n \hat{S}^z_i$ commutes with the
Heisenberg Hamiltonian $\hat{H}$, Eq. \eqref{eq:hamiltonian:heisenberg}.

Due to these fundamental differences in both sampling space and algorithmic
structure, a direct comparison of the two methods is not trivial, and
determining which approach is superior depends heavily on the specific context
and objectives of the simulation. In the following, the two methods will be
compared by looking at the number of {Monte Carlo steps} $N$ required to
achieve the same accuracy for the variational energy. The accuracy is quantified
by the relative error of the variational energy which is defined as
\begin{equation}
    \varepsilon (N) = \left|\frac{E(N) - E_{\mathrm{b}}}{E_{\mathrm{b}}}\right|
    \label{eq:relative_error}
\end{equation}
where $E(N)$ is the variational energy calculated and $E_\mathrm{b}$ serves as a
baseline ground state energy.

{The latter is obtained using MH sampling by averaging over $32$ MC
chains of $100.000$ MH sweeps, which is defined as attempting to flip pairs of
antiparallel spins (MH flip) $n$ times, see Fig. \ref{fig:figuremethod}b.}

In order to determine the computational advantage of the sIM, we need to
determine the number of steps $N$ required for iso-accuracy for the same
pre-trained model which is defined as
\begin{equation}
    \varepsilon_\mathrm{\text{sIM}} (N_\mathrm{sIM}) = \varepsilon_\mathrm{\text{MH}} (N_\mathrm{MH}) \\
    \label{eq:isoaccuracy}
\end{equation}
{where the subscripts sIM and MH indicate the values of $N$ and
$\varepsilon$ for sIM and MH sampling, respectively.}

{The computational advantage is identified when the total runtime of sIM
sampling is less than that of MH sampling,}
\begin{equation}
    N_\mathrm{sIM} T_\mathrm{sIM} < N_\mathrm{MH} T_\mathrm{MH}.
    \label{eq:computationaladvantage}
\end{equation}
{where $T_\mathrm{sIM} (T_\mathrm{MH})$ denotes the runtime per step of sIM (MH)
step.}

To obtain the number of steps for iso-accuracy, we will now focus on deriving an
explicit formula to compare sampling algorithms. {Since both the sIM and
MH are Markov chain Monte Carlo (MCMC) methods, they generate correlated
samples. For this case, the sampling error for an observable $O$ is given by
\begin{equation}
    \sigma_{\overline{O}}(N, \rho) = \sqrt{\frac{1 + 2 \sum_{l=1}^{N} \rho(l)}{N} \mathrm{Var}(O)}
    \label{theory:eq:montecarlo:erroranalysiswithautocorr}
\end{equation}
in the limit of large $N$, where $\rho(l)$ is the autocorrelation function and
$\mathrm{Var}(O)$ is the variance of observable $O$. The autocorrelation
function decays exponentially with the autocorrelation time $\tau$: for large
$l$, $\rho(l) \sim e^{-l/\tau}$ \cite{sokal_montecarlomethods_1997,
newman_montecarlomethods_1999, muller-krumbhaar_dynamicpropertiesmonte_1973}.}

\begin{figure}
\centering
\includegraphics{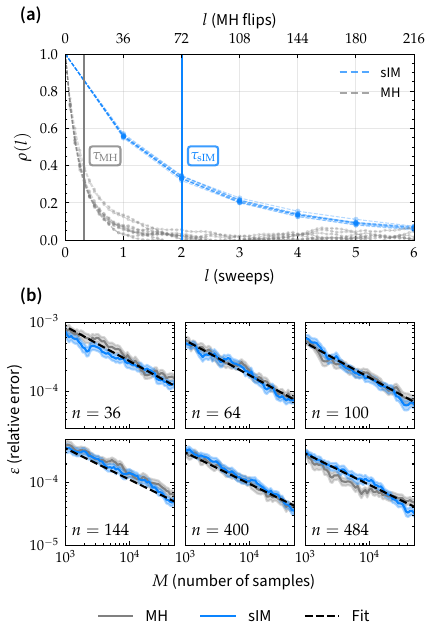}
\caption{\label{fig:erelscaling} {\textbf{(a)} Autocorrelation function
as a function of $l$ for $n = 36$, $\alpha = 2$. Bottom axis shows the interval
in units of the sweep size. The top axis gives the number of MH flips, which
only applies to MH sampling. The measured autocorrelation time is
$\tau_\mathrm{MH} = 0.32 \pm 0.02$ MH sweeps and $\tau_\mathrm{sIM} = 2.01 \pm
0.01$ sIM sweeps. On the top x-axis, the MH sweeps is converted into MH flips.
\textbf{(b)} The relative error, $\varepsilon(M)$ as a function of $M$ for $n =
36, 64, 100, 144, 400, 484$ spins at $\alpha = 2$. Here $M$ corresponds to the
number of samples at a sampling interval of $\Delta t_\mathrm{sIM,MH} =
\tau_\mathrm{sIM, MH}$ respectively. The relative error for sIM and MH sampling
is calculated as a mean value over $32$ runs and the error of the relative error
indicate the standard error of the mean. The fit highlights the scaling with
$M^{-1/2}$. The shaded error bars signify one standard deviation estimated from
$32$ chains.}}
\end{figure}

{This behavior is shown in Fig. \ref{fig:erelscaling}a for both MH and
sIM sampling as a function of the number of sweeps (for sIM and MH) and also in
flips (for MH). Here a sIM sweep is defined as performing sIM updates until a
new configuration with zero net magnetization is obtained. A sIM update consists
of updating all hidden and then all visible spins using chromatic Gibbs
sampling. }This {resembles} the hardware realization of a sIM, where
bits flip independently with probabilities given by the applied input. Unlike MH
sampling, a sIM cannot be constrained to a fixed magnetization sector without
distorting the desired Boltzmann distribution. {Since observables are
only defined for samples with zero magnetization, we discard samples with
non-zero magnetization.}

{Due to the correlated nature of MCMC, it is more efficient in practice
to evaluate observables at a sampling interval $\Delta t = N/M > 1$, where $M$
is the number of samples used in the calculations. For large $M$ and by scaling
$\tau$ in units of $\Delta t$, we can approximate $\sum_{k=1}^{M}
\rho(k) \approx \sum_{k=1}^{M} e^{-k / (\Delta t/\tau)}
\approx 1/(e^{\Delta t/\tau} - 1)$.}

The relative error $\varepsilon({M})$ scales with the sampling error of
the variational energy given by Eq.
\eqref{theory:eq:montecarlo:erroranalysiswithautocorr}, which decreases when
increasing the number of {samples} until the error becomes on the same
order of the baseline. In the case of $\sigma_E \gg \sigma_{E_b}$ {and
large $M$}, the relative error is then given by
\begin{equation}
    \varepsilon (M) \sim \sigma_{\overline{E}} (M, \Delta t/\tau) = \sqrt{\frac{1+ 2/(e^{\Delta t/\tau} - 1)}{M} \mathrm{Var}(E)}.
    \label{eq:erelscaling}
\end{equation}

{In Fig. 2b, sampling with $\Delta t = \tau$ for sIM and MH gives the
same scaling for the relative error which is confirmed by the fitted black
dashed line $\varepsilon_\mathrm{fit}(M) = a M^{-1/2}$. This confirms the
scaling in Eq. \eqref{eq:erelscaling} and that shows sampling at the fixed
interval $\Delta t=\tau$ and given number of samples $M$ yields the same
relative error for both methods.}

{To estimate the computational cost, we want to know the number of steps
$N$ required to reach iso-accuracy. For that we want to take the same number of
samples from both samplers and we set $M_\mathrm{sIM} = M_\mathrm{MH}$ and
substitute this into the iso-accuracy condition, Eq. \eqref{eq:isoaccuracy},
which yields}
\begin{equation}
    \frac{\tau_\mathrm{sIM}}{\tau_\mathrm{MH}} = \frac{\Delta t_\mathrm{sIM}}{\Delta t_\mathrm{MH}}
    \label{eq:ratiostepsmethod3}
\end{equation}
where we used that $\mathrm{Var}(E)$ is the same for both methods. Hence, the
sampling interval and therefore the number of steps needed for iso-accuracy is
determined by the autocorrelation time. Interestingly, for the sIM sampling the
number of steps will be related to the sampling efficiency of the Ising model
onto which the NQS is mapped.

In the following, we directly use Eq. \eqref{eq:ratiostepsmethod3} to estimate
the number of steps required on a sIM as compared to MH sampling by measuring
the autocorrelation of sIM and MH sampling. To this end, we consider pre-trained
RBM models which are optimised for the ground state of the 2D ($L \times L$)
antiferromagnetic Heisenberg model with stochastic reconfiguration
\cite{carleo_solvingquantummanybody_2017, sorella_greenfunctionmonte_1998}, see
Appendix \ref{appendix:trainingparameters}. {We also compared the
accuracy to QMC \cite{sandvik_finitesizescalinggroundstate_1997} in Appendix
\ref{appendix:accuracy}.} Each pre-trained RBM model has a corresponding Ising
model, with a given exchange matrix and bias vector, Eq.
\eqref{ising:equation:Jmatrix} and \eqref{ising:equation:biasvector}
respectively. We constructed a data set of several these pre-trained models with
various combinations of $\alpha$ and $\nspins$ ($= L^2$). This enables a
systematic analysis of the scaling advantages of sIM sampling as function of the
representational power and system size. For each combination of $\alpha$ and
$n$, $5$ models were trained and for each model the sIM and MH autocorrelation
time was measured.
\begin{figure}
\includegraphics{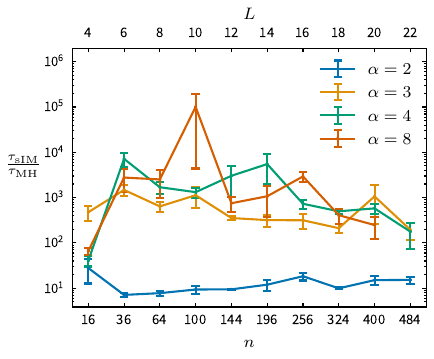}%
\caption{\label{fig:ratioautocorrelationtime} Ratio of autocorrelation time
$\tau_\mathrm{sIM} / \tau_{MH} = \Delta t_\mathrm{sIM}/ \Delta t_{MH}$ for
different models. Each point is the average of $5$ pre-trained models with the
same $\alpha$ and $n = L^2$.}
\end{figure}
The results at $\alpha = 1$ were excluded from the figure because they are less
accurate
\cite{carleo_solvingquantummanybody_2017,fabiani_investigatingultrafastquantum_2019},
moreover extensive hyperparameter tuning is required to consistently find an
accurate ground state for this value. The autocorrelation time per model was
measured for $5$ ($10$) chains for sIM (MH) sampling, see Appendix
\ref{appendix:autocorrelation}.

The results for the ratio {$\tau_\mathrm{sIM}/\tau_\mathrm{MH}$} are
shown in Fig \ref{fig:ratioautocorrelationtime}. It is clear that the ratio of
the autocorrelation time of sIM and MH can differ by orders of magnitude.
Interestingly, models with $\alpha = 2$ have a significantly lower
autocorrelation time compared to the models with $\alpha > 3$, which we
elaborate on in the discussion below.

\begin{figure}
    \includegraphics{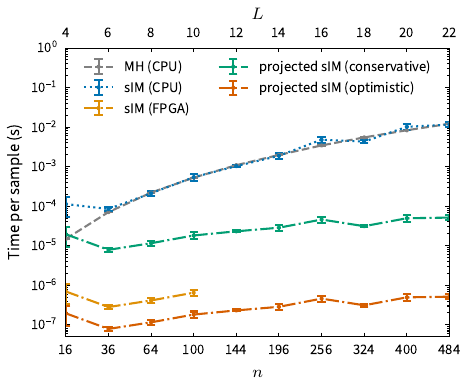}
    \caption{\label{fig:projection} Performance comparison for generating
    {one sample} across five methods as a function of $\nspins = L^2$
    for $\alpha = 2$. The baseline, MH (CPU) obtained using UltraFast
    \cite{berns_ultrafastjl_2025}, is a traditional NQS implementation with MH
    sampling. Furthermore, there are four sIM implementations: CPU-based
    implementation, FPGA running at 70 MHz
    \cite{patel_isingmodeloptimization_2020}, conservative projection based on
    \cite{gallo_64coremixedsignalinmemory_2023}, and optimistic projection based
    on \cite{sutton_autonomousprobabilisticcoprocessing_2020}.}
\end{figure}

Equipped with the ratio of the autocorrelation time, we focus on projecting the
possible computational advantage of deployment on sIMs. To this end, we focus on
the $\alpha=2$ case and consider reported values of the runtime per step on
physical sIMs. 
{Since NQS literature often reports the number of required samples at a sampling interval of one MH sweep, we will report the time to generate the same effective sample for MH and sIM. The time per sample on the sIM is calculated from the ratio of the sampling intervals, Eq. \eqref{eq:ratiostepsmethod3}, with $\Delta
t_\mathrm{MH} = 1 \mathrm{MH \ sweep}$ and the runtime per update on the sIM hardware:
\begin{equation}
    \text{Time per sample} = \Delta t_\mathrm{sIM} T_\mathrm{sIM} = \frac{\tau_\mathrm{sIM}}{\tau_\mathrm{MH}f}T_\mathrm{sIM}^\mathrm{update}
\end{equation}
where $\tau_\mathrm{sIM}$, $\tau_\mathrm{MH}$, $f$ are all dependent on the
number of spins $n$ and $\alpha$. Since the sIM is not constrained to the
magnetization zero sector, the fraction $f$ enters, which is the ratio of the
number of samples with zero magnetization to the total number of samples:
$T_\mathrm{sIM} = T_\mathrm{sIM}^\mathrm{update} / f$. The MH sampling baseline
on CPU, $\text{Time per sample} = T_\mathrm{MH}$, is directly measured in our
experiments and is also dependent on $n$ and $\alpha$. On all (hardware)
implementations, a single MC chain is assumed.}

In Fig. \ref{fig:projection}, the MH (CPU) baseline, is the performance of
UltraFast \cite{berns_ultrafastjl_2025} on a machine with dual AMD
EPYC 7601 CPU. The sIM (CPU) corresponds to a chromatic Gibbs sampling code
written in Julia on the same machine. This implementation uses multithreading
for matrix-vector multiplication \cite{berns_predictingsamplingadvantage_2025_code}. Benchmarks for a (single-chain) sIM GPU
implementation on an NVIDIA L40S GPU are omitted, as it only outperforms the sIM
CPU implementation for systems with $\nspins > 700$ spins, which is larger than
the system sizes considered here, see Appendix \ref{appendix:gpusim}.

The sIM FPGA estimate \cite{patel_logicallysynthesizedhardwareaccelerated_2020a}
features a chromatic Gibbs sampling algorithm for the RBM and shows problem
sizes up to $200$ visible spins which are fully connected to $200$ hidden spins.
The samples are generated at $70$ MHz, i.e. $14.3$ ns per sample.

The conservative sIM projection is based on a device combining the IBM HERMES
Project Chip with stochastic MTJs \cite{gallo_64coremixedsignalinmemory_2023}.
This chip design features 64 analog crossbars of size $256 \times 256$ for MVMs
with a latency of 133 ns \cite{gallo_64coremixedsignalinmemory_2023}.
{We assume that the} typical latency of the stochastic MTJs is {
$2$} ns
\cite{schnitzspan_nanosecondtruerandomnumbergeneration_2023,sutton_autonomousprobabilisticcoprocessing_2020}.
{Implementing chromatic Gibbs sampling requires two MVMs, splitting the
MVM over multiple cores (if MVM is bigger than one crossbar) and two read-outs
of the MTJs. Therefore, we estimate 400 ns per sIM sweep for this device, see
Appendix \ref{appendix:estimatingsimperformance} for more details.}

The optimistic sIM projection follows the analysis of
\cite{sutton_autonomousprobabilisticcoprocessing_2020} for a projected
asynchronous sIM which is based on a $100 \times 100$ crossbar with a latency of
$10$ ps \cite{gu_technologicalexplorationrram_2015}. Again combining this with
the stochastic MTJs, we estimate a latency of $4$ ns per sIM sweep, or 2 orders
of magnitude faster than the conservative projection, {see Appendix
\ref{appendix:estimatingsimperformance}.}

From Fig. \ref{fig:projection}, it is clear that the CPU based sIM
implementation performs similar to the MH (CPU, UltraFast
\cite{berns_ultrafastjl_2025}) implementation. The conservative
projection is 2 orders of magnitude faster than the baseline and demonstrates a
more favorable scaling with system size. Interestingly, the FPGA implementation
is faster than conservative projection. Lastly, the optimistic projection is 2
orders of magnitude faster than the conservative projection and slightly faster
than the FPGA implementation.

\section{Discussion}

To better understand the increased autocorrelation at $\alpha > 2$ as compared
to $\alpha=2$, we studied the differences between the probabilistic flipping of
visible and hidden spins. This is of key relevance since we are considering the
variational quantum energy as key observable, which depends only on visible
spins. The increased autocorrelation time, which is reflected in the freezing of
the visible spins, can be explained by the height of the energy barrier required
to flip a visible spin. The energy barrier is defined as $\Delta {E_i} =
H_\text{ising}({z}_i \rightarrow -{z}_i) -
H_\text{ising}({z})$ and the height is determined by the coupling
strength of a visible spin to all the ${m} = \alpha n$ hidden spins. In
Appendix \ref{appendix:autocorrelationanalysis}, we show that the mean coupling
strength decreases at a slower rate than $\alpha$ increases. Combined with the
fact that connectivity per visible spin increases with $\alpha$, gives a net
growth of the energy barrier and therefore exponential suppression of flipping
visible spins. We further confirm this interpretation by observing a
significantly higher rate of probabilistic flipping of the hidden spins,
consistent with their connectivity remaining constant as $\alpha$ increases.
Hence, the hidden and visible spins feature very much distinct dynamics. This
explanation suggests great potential for using sparse DBMs, which may suffer
much less from the high energy barriers, while still offering great
expressiveness \cite{niazi_trainingdeepboltzmann_2024}.

Based on the ratio of the autocorrelation time, we projected the performance of
sampling on a sIM for $\alpha = 2$ and compared it to the MH sampling
traditionally used for NQS. It must be noted that hardware implementations
require challenges to be overcome, such as quantization of the weights and
biases, resistance to noise, and possibly reducing connectivity. Next to that,
in order to treat the two methods on the same footing, we compare single chain
MH sampling with single chain sIM sampling. This is done because we are
interested in optimizing performance of a single chain and splitting into
multiple chains for hardware implementations can be done by using more devices.
Lastly, for the considered system sizes the CPU sIM implementation is faster
than the GPU implementation. For MH sampling, only a CPU implementation was
benchmarked, we expect that for larger system sizes some performance gain can be
expected using an optimized GPU implementation.

In addition to the runtime estimates, we can also make a back of the envelope
energy comparison. For the conservative projection, we use the largest system
size ($\nspins = 484$, $\alpha=2$) and that the energy usage per AIMC core is
$100$ mW \cite{gallo_64coremixedsignalinmemory_2023}, 4 cores are used and the
device requires 400 ns per step. For MH (UltraFast) on an AMD EPYC 7601 CPU, we
use the reported $180$ W TDP divided by the $32$ cores and measured average
runtime per core while utilizing $16$ cores. With these settings, the
conservative projection yields about $3$ orders of magnitude higher energy
efficiency than MH (UltraFast) on CPU. {The TDP gives a rough estimate
of the energy usage, in future work, more accurate energy comparisons can be
obtained by directly measuring the energy, which is possible using the energy
aware runtime (EAR) \cite{julitacorbalan_energyawareruntime_}}.

We further note that the sampling advantages can be model specific. For example,
already for Heisenberg spin ladders, it was found that the sampling complexity
is higher than the square lattice case \cite{hofmann_rolestochasticnoise_2022},
as well as several other models such as the highly frustrated $J_1$-$J_2$ models
\cite{carleo_constructingexactrepresentations_2018,nomura_purifyingdeepboltzmann_2021}.

In summary, we have shown that the sampling advantage of sIMs for NQS is
determined by the sampling efficiency of the Ising model onto which the problem
is mapped. The sampling efficiency can be quantified hardware-agnostic by the
autocorrelation time. Projected hardware can accelerate the sampling for shallow
networks by 2 to 4 orders of magnitude. Wider networks, i.e. higher $\alpha$,
suffer from high energy barriers which increase autocorrelation time several
orders of magnitude and may be tamed by considering sparse models.

Given the highly promising sampling advantages, we expect that our studies open
new opportunities for large scale quantum simulations. Particularly interesting
next steps include investigation of sampling advantages far away from the ground
states, which is relevant for training NQS. Further changing the topology of the
Ising model to models such as {sparse DBMs
\cite{niazi_trainingdeepboltzmann_2024} is very interesting. DBMs have the same
blocking structure as RBMs and hence our methodology can straightforwardly be
applied to such models as well and potentially the sparsity can be used to tune
the autocorrelation time without sacrificing accuracy}. Another direction is to
study frustrated models as well as unitary quantum dynamics on the sIM, which
would involve using an ansatz with amplitude and phase. {Beyond
applications to NQS, the methodology developed in this paper can
straightforwardly be used for estimating possible speedups of hardware
improvements on Boltzmann machine simulation of general machine learning
workloads, which is particularly interesting in view of the rapid development of
Ising machine hardware.}

\begin{acknowledgments}
This publication is part of the project NL-ECO: Netherlands Initiative for
Energy-Efficient Computing (with project number NWA. 1389.20.140) of the NWA
research programme Research Along Routes by Consortia which is financed by the
Dutch Research Council (NWO). JHM acknowledges funding from the VIDI project no.
223.157 (CHASEMAG) and KIC project no. 22016 which are (partly) financed by the
Dutch Research Council (NWO), as well as support from the European Union Horizon
2020 and innovation program under the European Research Council ERC Grant
Agreement No. 856538 (3D-MAGiC) and the Horizon Europe project no. 101070290
(NIMFEIA). GF and DR acknowledge the support the project PRIN 2020LWPKH7 - The
Italian factory of micromagnetic modeling and spintronics funded by the Italian
Ministry of University and Research (MUR). DR acknowledges the support from the
project PE0000021, ”Network 4 Energy Sustainable Transition—NEST”, funded by the
European Union—NextGenerationEU, under the National Recovery and Resilience Plan
(NRRP), Mission 4 Component 2 Investment 1.3—Call for Tender No. 1561 dated
11.10.2022 of the Italian MUR (CUP C93C22005230007); project D.M. 10/08/2021 n.
1062 (PON Ricerca e Innovazione), funded by the Italian MUR.

\end{acknowledgments}

\nocite{berns_data_sampling_advantage_2026}

\newpage

\appendix

\section{Alternative mappings}
Note that there are alternative mappings from NQS to an Ising Model. In
\cite{chowdhury_machinelearningquantum_2023}, mapping from NQS to an Ising model
was demonstrated for the first time. Another method is described in the
supplementary material of \cite{nomura_purifyingdeepboltzmann_2021}.

\section{Neural Quantum States}
\subsection{Training parameters}
\label{appendix:trainingparameters}
For the training of the models, stochastic reconfiguration (SR)
\cite{carleo_solvingquantummanybody_2017,sorella_greenfunctionmonte_1998} was
used. The gradient descent update rule of SR reads
\begin{equation}
    \mathcal{W}_k (p+1) = \mathcal{W}_k (p) - \eta(p) S_{k k^\prime}^{-1} \mathcal{F} (\mathcal{W}_k (p))
    \label{eq:vmc:stochastic_reconfiguration}
\end{equation}
where $p$ is the training iteration and $S_{k k^\prime}$ is the S-matrix given
by
\begin{equation}
    S_{k k^\prime} = \langle O_k^{\star} O_{k^\prime} \rangle - \langle O_k^{\star} \rangle \langle O_{k^\prime} \rangle
    \label{eq:vmc:smatrix}
\end{equation}
and $\mathcal{F}_k$ is given by
\begin{equation}
    \mathcal{F}_k = \langle E_\text{loc} O^{\star}_k \rangle - \langle E_\text{loc} \rangle \langle O^\star_k \rangle
    \label{eq:vmc:fk}
\end{equation}
with $O_k(s)$ defined as
\begin{equation}
    O_k(s) = \frac{1}{\psi_{\mathcal{W}}(s)} \frac{\partial \psi_{\mathcal{W}}(s)}{\partial \mathcal{W}_k} = \frac{\partial}{\partial \mathcal{W}_k} \log \psi_\mathcal{W} (s).
    \label{eq:vmc:gradients}
\end{equation}

Metropolis-Hastings (MH) sampling was used for computing expectation values. The
thermalization time was configured to $200$ MH sweeps and samples were taken one
sweep apart. The number of training iterations was between $300$ and $600$
iterations and the learning rate $\eta = 0.005$. For the sampling settings and
regularization two different sets of parameters were used depending on the
number of spins and $\alpha$, which is given in Table
\ref{tab:nspins_alpha_lowhigh}. There is a set with 'low' which corresponds to
$2000$ MC samples and 'high' corresponds to $10000$ MC samples. Also the
regularization is different for these two sets and is given by:
\begin{equation}
    S = S + \epsilon(p) I
\end{equation}
where $\epsilon(p) =  \max(\epsilon, \epsilon_0 \cdot b^p)$. For low, $\epsilon
= 0.0001$, $b = 0.9$ and $\epsilon_0 = 100$ and for high $\epsilon = 0.001$, $b
= 0.85$ and $\epsilon_0 = 10$.

\begin{table}[h]
    \centering
    \caption{Summary of hyperparameters depending on ${n}$ and $\alpha$ for training with standard and modified RBM ansatze.}
    \label{tab:nspins_alpha_lowhigh}
    \begin{tabular}{@{}cccccc@{}}
                  & \multicolumn{5}{c}{$\alpha$}    \\ \cmidrule(l){2-6} 
        ${n}$ & $1$ & $2$  & $3$  & $4$  & $8$  \\ \midrule
        $16$      & low & low  & low  & low  & low  \\
        $36$      & low & low  & low  & low  & low  \\
        $64$      & low & low  & low  & low  & low  \\
        $100$     & low & low  & low  & low  & low  \\
        $144$     &     & low  & low  & low  & high \\
        $196$     &     & low  & low  & low  & high \\
        $256$     &     & high & high & high & high \\
        $324$     &     & high & high & high & high \\
        $400$     &     & high & high & high & high \\
        $484$     &     & high & high & high &      \\ \bottomrule
        \end{tabular}
    
\end{table}

\subsection{{Accuracy of pre-trained models}}
\label{appendix:accuracy}

\begin{figure*}[htb!]
\center
\includegraphics{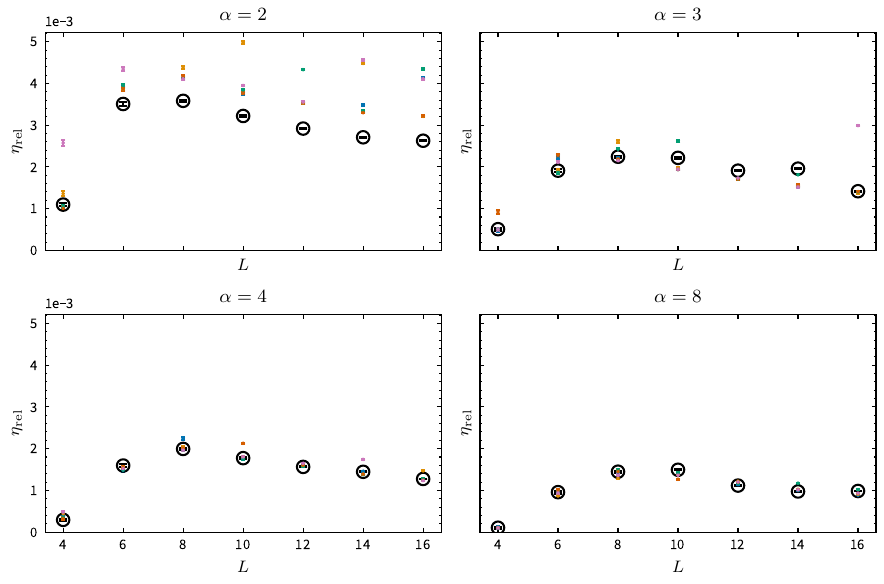}
\caption{The relative error, Eq. \eqref{sup:eq:erel:compnqsqmc}, as a function
of the side length $L$ of 2D antiferromagnetic Heisenberg model. The QMC ground
state energies are taken from \cite{sandvik_finitesizescalinggroundstate_1997}
for comparison. Colored symbols refer to the results obtained from 5 different
pretrained models per $\alpha$ and $\nspins$. Black circles with error bars are
results obtained with ansatz used in
\cite{carleo_solvingquantummanybody_2017,fabiani_investigatingultrafastquantum_2019}.
\label{fig:erel_eloc}}
\end{figure*}

{The accuracy of the ground state energy of the pre-trained RBM ansatz
is compared to the QMC ground state energy from}
\cite{sandvik_finitesizescalinggroundstate_1997} {in Fig.
\ref{fig:erel_eloc}.} The accuracy is expressed using the relative error given
by
\begin{equation}
    \eta_\mathrm{rel} = \left|\frac{E_\mathrm{NQS}-E_\mathrm{QMC}}{E_\mathrm{QMC}}\right|
    \label{sup:eq:erel:compnqsqmc}
\end{equation}
where $E_\mathrm{NQS}$ is the NQS ground state and $E_\mathrm{QMC}$ is the QMC
ground state from \cite{sandvik_finitesizescalinggroundstate_1997}.

{Furthermore,} the accuracy of the {pre-trained models with the
wave function ansatz given in} Eq. \eqref{ising:equation:modifiedansatz}, are
compared to the 'standard' RBM ansatz used in
\cite{carleo_solvingquantummanybody_2017,
fabiani_investigatingultrafastquantum_2019} in Fig. \ref{fig:erel_eloc}. For all
the models, $32$ chains of $10000$ MH sweeps were used. {For some of
these models the Markov chain got stuck in samples with high $E_\mathrm{loc}$,
we excluded these chains based on the average quantum energy: $E > -0.55$, which
is much higher than results from e.g. Linear Spin Wave Theory $E = -0.658$ per
spin \cite{fazekas_lecturenoteselectron_1999}. The relative error was then
calculated over the remaining chains}. The models where chains are excluded are
listed in the MH columns of Table \ref{tab:excludedmodels}.

\section{Autocorrelation time}
\label{appendix:autocorrelation}

\begin{figure}
\centering
\includegraphics{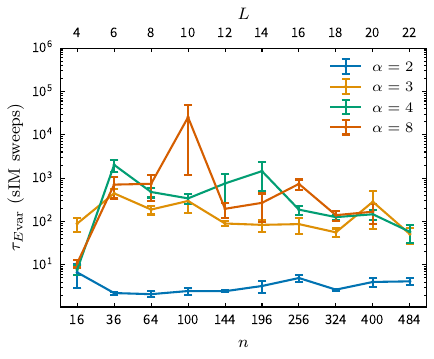}
\caption{\label{fig:autocorrelation_Ising} Autocorrelation time of the
variational energy for sIM sampling. For every combination of $\alpha$ and
$\nspins$ there is averaging over $5$ pre-trained models.}

\end{figure}

\begin{figure}
\centering
\includegraphics{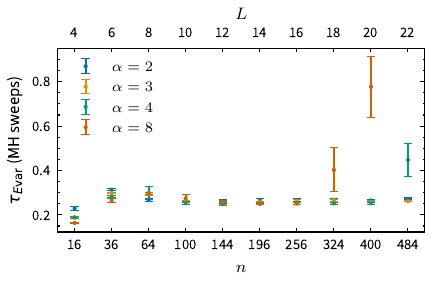}
\caption{\label{fig:autocorrelation_UltraFast} Autocorrelation time of the
variational energy for the Metropolis-Hastings sampling. For every combination
of $\alpha$ and $\nspins$ there is averaging over $5$ pre-trained models.}

\end{figure}

{Using Eq. \eqref{eq:ratiostepsmethod3}, we can estimate the sampling
interval from the autocorrelation time and therefore the number of steps
required on a sIM to reach iso-accuracy compared to MH sampling. To find the
autocorrelation time from the autocorrelation function $\rho(j)$, we use the
integrated autocorrelation time approach, where the autocorrelation function is
integrated up to a cut-off value $C$:
\begin{equation}
    \tau_\mathrm{int} = \frac{1}{2} + \sum_{j=1}^{C} \rho(j).
    \label{eq:integrated_autocorrelation}
\end{equation}
The cutoff value is the smallest $C$ such that $C \geq 4 \tau_\mathrm{int} + 1$
and for that $C$ we have $\tau \approx \tau_\mathrm{int}$
\cite{joseph_markovchainmonte_2020}}. {In Fig.
\ref{fig:autocorrelation_Ising} (\ref{fig:autocorrelation_UltraFast}) the
measured autocorrelation time of sIM (MH) is shown expressed in sIM (MH) sweeps
respectively.}

The following sampling settings were used for calculating the autocorrelation
time. For the MH sampling, the sampling interval was set to 
\begin{equation}
    {\Delta t} = \max(0.01 \cdot \nspins, 1) \ \text{MH flips}
\end{equation}
and $10$ chains of each $32768$ samples were taken.

For sIM, the sampling interval was taken at a variable interval. The interval
was chosen such that a chain contained at least $1500$ autocorrelation times and
the autocorrelation time measured in the samples was at least larger than $2$
(except for $\alpha \leq 2$ where at least larger than $1.5$). There are at
least $5$ chains of at least $32768$ samples taken for each model.

{As mentioned in Appendix \ref{appendix:accuracy}, some models suffered
from Markov chains getting stuck. This also occured for measuring the
autocorrelation time of MH and sIM sampling. Here, instead of excluding chains,
we excluded the entire pre-trained model if any of the chains had an average
energy $E > -0.55$ for either sIM and/or MH sampling. The excluded models are
listed in Table \ref{tab:excludedmodels}}. Additionally, results for $\alpha =
1$ are already excluded due to poor performance.

Note, that for calculating the variational energy longer Markov chains were used
compared to autocorrelation time resulting in a higher probability of getting
stuck (e.g. you can only get stuck ones).

\begin{figure}
\centering
\includegraphics{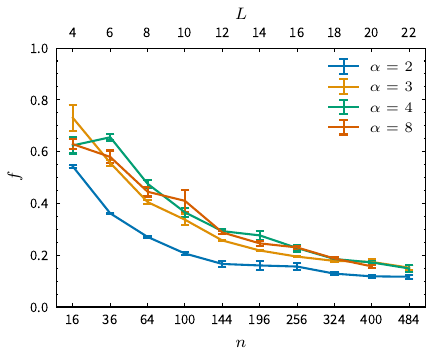}
\caption{\label{fig:mag0ratio} The ratio of samples in the magnetization 0
sector $f$ sampled by sIM sampling as a function of the number of spins $n$
and side length $L$ for different hidden layer densities $\alpha$.}

\end{figure}

\begin{table}
    \centering
    \caption{Models excluded from the analysis. Model Id=0...4 refer to labels
    used in the dataset for 5 different models per $n$ and $\alpha$. The
    criterion for exclusion is based on the error in the relative energy, see
    the discussion in Appendix \ref{appendix:autocorrelation}}
    \label{tab:excludedmodels}
    \begin{tabular}{ccccc}
        ${n}$ & $\alpha$ & Id & sIM & MH \\
        \midrule
        64  & 4  & 1  & x & x \\
        64  & 4  & 3  & x & x \\
        144 & 4  & 4  &  & x  \\
        {64}  & {8}  & {2}  & {x} &  \\
        64  & 8  & 3  & x &   \\
        100 & 8  & 1  &  & x  \\
        100 & 8  & 4  & x & x \\
        400 & 4  & 1  & x &   \\
        144 & 8  & 0  & x &   \\
        144 & 8  & 2  &  & x  \\
        196 & 8  & 3  &  & x  \\
    \end{tabular}
\end{table}

\section{{Blocking Gibbs sampling}}
\label{appendix:blockinggibbssampling}
{ In this article, we implement blocking Gibbs sampling which allows to
update all the hidden spins in parallel based on all the visible spins and vice
versa. The blocking Gibbs sampling can be implemented on hardware Ising Machines
and we have implemented it in Julia for the purpose of this article
\cite{berns_predictingsamplingadvantage_2025_code}.}

{
The blocking Gibbs sampling starts with a random initialization of the visible
spins $s$ and then iteratively updates the hidden spins $x$ and visible spins
$s$. The hidden spins are updated based on the input from the visible spins and
is given by the following equation:
\begin{align}
I_j &= \sum_{i=1}^{n} W_{ij} s_i + b_j \label{eq:appendix:synapsehidden} \\
x_j &= \mathrm{sgn} \left( \tanh \left( I_j \right) - r_j \right) \label{eq:appendix:neuronhidden}
\end{align}
where $r_j$ are $M$ uniform random numbers between $-1$ and $1$. The visible
spins are updated based on the input from the hidden spins and is given by the
following equation:
\begin{align}
I_i &= \sum_{j=1}^{M} W_{ij} x_j \label{eq:appendix:synapsevisible}\\
s_i &= \mathrm{sgn} \left( \tanh \left( I_i \right) - r_i \right) \label{eq:appendix:neuronvisible} 
\end{align}
where $r_i$ are $n$ uniform random numbers between $-1$ and $1$.}

\section{Performance sIM GPU}
\label{appendix:gpusim}

\begin{figure}
    \centering
    \includegraphics{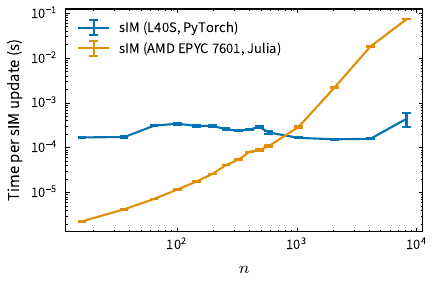}
    \caption{\label{fig:autocorrelation_GPU} Comparison of the time per sIM
    update for a single chain sIM GPU implementation and a single chain sIM CPU
    implementation as a function of the number of visible spins $\nspins$ for
    $\alpha = 2$. Both implementations use chromatic Gibbs sampling and a sIM
    update refers to one update of all visible spins and one update of all
    hidden spins. The GPU implementation is written in PyTorch and the CPU
    implementation is written in Julia.}
\end{figure}

The performance of a single chain sIM (chromatic Gibbs) GPU implementation
written in PyTorch on an NVIDIA L40S GPU was benchmarked for random weights and
biases \cite{berns_predictingsamplingadvantage_2025_code}. In Fig. \ref{fig:autocorrelation_GPU}, the performance of that GPU
implementation is compared to the single chain sIM (chromatic Gibbs) CPU
implementation written in Julia. The GPU implementation is only competitive for
larger system sizes ($\nspins > 700$) for $\alpha = 2$.

\section{{Estimating the performance of the (hardware) sIM}}
\label{appendix:estimatingsimperformance}

{
Implementing blocking Gibbs sampling using stochastic Magnetic Tunnel Junctions
(MTJs) implements Eq. \ref{eq:appendix:neuronhidden} and
\ref{eq:appendix:neuronvisible} with stochastic MTJs. The matrix multiplication
in Eq. \ref{eq:appendix:synapsehidden} and \ref{eq:appendix:synapsevisible} can
be implemented using a crossbar. For the conservative
\cite{gallo_64coremixedsignalinmemory_2023} and optimistic
\cite{sutton_autonomousprobabilisticcoprocessing_2020} projections we assume
that they implement blocking Gibbs sampling. Since the typical latency of
stochastic MTJs stretches from $0.1 - 10$ ns
\cite{sutton_autonomousprobabilisticcoprocessing_2020,schnitzspan_nanosecondtruerandomnumbergeneration_2023},
we assume $t_\text{MTJ} = 2$ ns for the MTJ latency for both the optimisitic and
conservative projection.}

{
The main difference between the conservative and optimistic projection is the
crossbar latency. The conservative projection uses a crossbar of $256 \times
256$ with a latency of $t_\text{c} = 133$ ns
\cite{gallo_64coremixedsignalinmemory_2023} and the optimistic projection uses a
crossbar of $100 \times 100$ with a latency of $t_\text{c} = 10$ ps
\cite{sutton_autonomousprobabilisticcoprocessing_2020}.}

{ Every sIM update consists of updating all visible and hidden spins,
which requires two crossbar operations, Eq. \eqref{eq:appendix:synapsehidden}
and \eqref{eq:appendix:synapsevisible}, and two read operations of the
stochastic MTJs, Eq. \ref{eq:appendix:neuronhidden} and
\ref{eq:appendix:neuronvisible}. For the conservative projection we get,
\begin{equation}
\begin{split}
T_\mathrm{sIM}^\mathrm{update} &= 2 \cdot T_\text{c} + 2 \cdot T_\text{MTJ} + T_\mathrm{acc} \\
&= 2 \cdot 133 \text{ ns} + 2 \cdot 2 \text{ ns} + 130 \text{ ns} \approx 400 \text{ ns}
\end{split}
\end{equation}
where \(T_\mathrm{sIM}^\mathrm{update}\) is the time per sIM update, \(T_\text{c}\) is the time
for a crossbar operation, \(T_\text{MTJ}\) is the time for reading the
stochastic MTJs, and \(T_\mathrm{acc}\) is the assumed time to accumulate the
results of the up to 64 crossbars on one chip and to round it off to $400$ ns
for this rough estimate. For the optimistic projection we get,
\begin{equation}
\begin{split}
T_\mathrm{sIM}^\mathrm{update} &= 2 \cdot T_\text{c} + 2 \cdot T_\text{MTJ} + T_\mathrm{acc} \\
    &= 2 \cdot 10 \text{ ps} + 2 \cdot 2 \text{ ns} \approx 4 \text{ ns}
\end{split}
\end{equation}
where \(T_\mathrm{acc}\) is assumed to be negligible since $T_\text{c} \ll
T_\text{MTJ}$.}

\section{{Comparing energy usage of hardware sIM and MH sampling}}
\label{appendix:energycomparisonsim}
{ The first step is to calculate the energy usage of one sIM update on
the hardware of the conservative projection. For $n_\mathrm{spins} = 484$ and
$\alpha = 2$, 4 cores of the chip are used, each core uses $100$ mW and $400$ ns
per step, resulting in
\begin{equation}
E_\mathrm{sIM} = 4 \cdot 100 \text{ mW} \cdot 400 \text{ ns} = 160 \text{ nJ}
\end{equation}}

{
The second step is to estimate the energy usage of one MH sweep on the CPU. We
use the reported 180 W TDP of the AMD EPYC 7601 CPU and divide by the 32 cores
and the $12$ ms measured average runtime per core while utilizing 16 cores.}

{
\begin{equation}
E_\mathrm{MH} = \frac{180 \text{ W}}{32} \cdot 12 \text{ ms} = 67.5 \text{ mJ}
\end{equation}}

{The sampling interval for MH sampling is set to $1$ MH sweep and the
sampling interval for sIM sampling is set based on the autocorrelation time
ratio using Eq. \eqref{eq:ratiostepsmethod3}. In this case, the ratio of energy
usage for obtaining one sample with MH compared to sIM is given by
\begin{equation}
\gamma = \frac{E_\mathrm{MH}}{E_\mathrm{sIM}}\frac{\Delta t_\mathrm{MH}}{\Delta t_\mathrm{sIM} f} = \frac{67.5 \text{ mJ}}{160 \text{ nJ}} \frac{1}{129} \approx 3 \cdot 10^3
\end{equation}
where we used that $\Delta t_\mathrm{sIM} f = 129$ sIM updates for $n = 484$ and
$\alpha = 2$. Therefore, the energy usage of sIM sampling is more than 3 orders
of magnitude lower than MH sampling at $n = 484$ and $\alpha = 2$. }

\section{Energy barrier analysis}
\label{appendix:autocorrelationanalysis}

\begin{figure*}
    \centering
\includegraphics{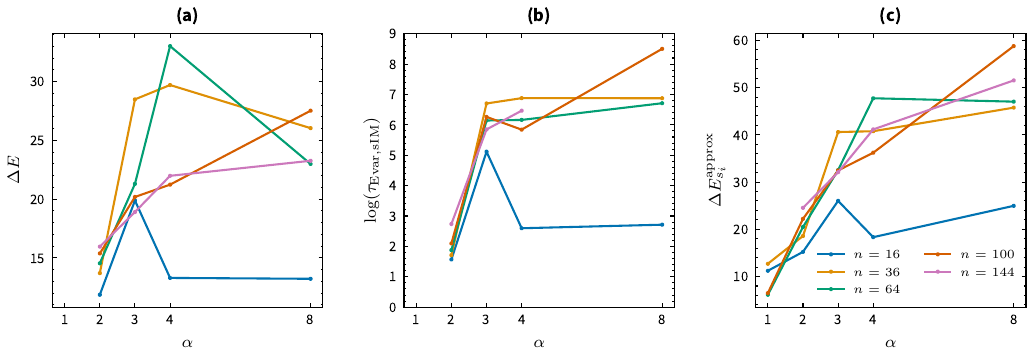}
\caption{\label{fig:appendix:deltaEWsum} \textbf{(a)} The average energy barrier
associated with flipping visible spins of sIM samples, given by $\Delta E =
\frac{1}{n}\sum_{i=1}^n \frac{1}{{M}} \sum_{\mu=1}^{{M}} \Delta E_i
({z}^\mu) = \frac{1}{n{M}} \sum_{\mu=1}^{{M}} \sum_{i=1}^n
E({z}_i^\mu \rightarrow -{z}_i^\mu) - E({z}^\mu)$.
\textbf{(b)} The logarithm of the autocorrelation time $\tau_\mathrm{Evar,
sIM}$. \textbf{(c)} The approximated energy barrier over the weight matrix
${\Delta E^\mathrm{approx}_{s_i}} = \frac{1}{n} \sum_{i=1}^n
\sum_{j=1}^{{m}} |W_{ij}|$ as a function of $\nspins$ and $\alpha$.
\textbf{(a, b, c)} The models considered have Id = 0 and the results are
presented as a function of $\alpha$ and $\nspins$.}

\end{figure*}

The energy barrier $\Delta E_i$ for flipping a spin in the Ising model is
defined as
\begin{equation}
    \Delta E_i ({z}) = H_\text{ising}({z}_i \rightarrow -{z}_i) - H_\text{ising}({z})
    \label{appendix:ising:equation:energybarrier}
\end{equation}
where $E({z})$ is the energy of the state ${z}$ given by Eq.
\eqref{ising:equation:isinghamiltonian}. The high autocorrelation time for
models with higher $\alpha$ is related to the freezing of the visible spins
which suggests that there is a high energy barrier for flipping the visible
spins. In Fig. \ref{fig:appendix:deltaEWsum}a, the average measured energy
barrier for flipping visible spins of typical samples is shown as a function of
$\nspins$ and $\alpha$. The energy barrier can be compared to the logarithm of
the autocorrelation time shown in Fig. \ref{fig:appendix:deltaEWsum}b, which
shows qualitatively the same behavior. This suggests that the autocorrelation
time is related to the energy barrier of the visible spins. Additionally, the
energy barrier of the visible spins can be roughly approximated. For that we
first simplify Eq. \eqref{appendix:ising:equation:energybarrier} to
\begin{equation}
    \Delta E_i ({z}) = 2 {z}_i \sum_{j=1}^{{m}} J_{ij} {z}_j + h_i {z}_i
    \label{appendix:ising:equation:energybarrierapprox}
\end{equation}
and extracting $\Delta E_i$ for the visible spins gives 
\begin{equation}
    \Delta E_{s_i}(x,s) = 2 s_i \sum_{j=1}^{{m}} W_{ij} x_j
    \label{appendix:ising:equation:energybarrierapproxvisible}.
\end{equation}
Note that $W_{ij}$ are the weights obtained from the ground state optimization
of the 2D Heisenberg model. This energy barrier can be roughly approximated by
setting $x_j = 1$, $s_i = 1$ and taking the absolute value of $W_{ij}$ which
gives
\begin{equation}
    {\Delta E^\mathrm{approx}_{s_i}} \sim \sum_{j=1}^{{m}} |W_{ij}|.
\end{equation}
This approximation is shown in Fig. \ref{fig:appendix:deltaEWsum}c, which also
shows qualitatively similar behavior as the autocorrelation time and the energy
barrier.

\begin{figure}

    \centering
\includegraphics{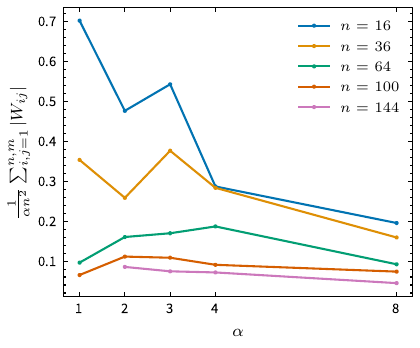}
\caption{\label{fig:appendix:meanW} The mean connection strength of the weight
matrix, given by $\frac{1}{\alpha n^2} \sum_{i,j=1}^{n,{m}} |W_{ij}|$ as
a function of $\nspins$ and $\alpha$}
\end{figure}

Lastly, Fig. \ref{fig:appendix:meanW} shows the mean connection strength of the
weight matrix
\begin{equation}
    \Delta E = \frac{1}{n} \sum_{i=1}^n \Delta E_i = \frac{1}{n} \sum_{i=1}^n \frac{1}{\alpha n} \sum_{j=1}^{{m}} |W_{ij}|.
    \label{appendix:ising:equation:meanconnectionstrength}
\end{equation}

\end{document}